\documentclass[preprint,preprintnumbers,amsmath,amssymb]{revtex4}
\usepackage{graphicx}
\usepackage{epstopdf}
\usepackage{dcolumn}
\usepackage{bm}
\usepackage{color}
 \usepackage{array,multirow,graphicx}

\begin{document}

\title{Dirac fermions in half-metallic ferromagnetic mixed Cr$_{1\textrm{-}x}$M$_x$PSe$_3$ monolayers}

\author{Juntao Yang,$^{1,2}$ Yong Zhou,$^{1}$ Yuriy Dedkov,$^{1,}$\footnote{E-mail: dedkov@shu.edu.cn} and Elena Voloshina$^{1,}$\footnote{E-mail: voloshina@shu.edu.cn}}

\affiliation{$^1$Department of Physics, Shanghai University, 99 Shangda Road, 200444 Shanghai, China}
\affiliation{$^2$School of Science, Hubei University of Automotive Technology, 167 Checheng West Road, Shiyan City, 442002 Hubei, China}

\date{\today}

\begin{abstract}
The electronic and magnetic properties of pristine CrPSe$_3$ and mixed Cr$_{1\textrm{-}x}$M$_x$PSe$_3$ (M = Zn, Cd, Hg) monolayers were studied using density functional theory including an on-site Coulomb term (DFT$+U$) and tight-binding approach (TBA). While pristine CrPSe$_3$ monolayer has an antiferromagnetic (AFM) ground state, its alloying with MPSe$_3$ may give rise to half-metallic ferromagnet (HMF) with high Curie temperature. The resulting monolayers demonstrate single-spin Dirac cones of mainly Cr-$d$ character located in the first Brillouin zone directly at the Fermi energy. The calculated Fermi velocities of Dirac fermions indicate very high mobility in mixed Cr$_{1\textrm{-}x}$M$_x$PSe$_3$ monolayers, that makes this material appealing for low-dimensional spintronics applications.
\end{abstract}

\maketitle

Discovery of the fascinating transport properties of graphene~\cite{Novoselov:2005es,Zhang:2005gp,Geim:2007hy} stimulated the enormous interest to the family of two-dimensional (2D) materials, leading to the success in the synthesis or exfoliation from bulk of different 2D materials, like $h$-BN~\cite{Oshima:1997ek,Tonkikh:2016ck,Zhang:2017jj}, black phosphorene~\cite{Carvalho:2016cv}, silicene~\cite{Houssa:2015ez,Molle:2018jl}, transition-metal dichalcogenides~\cite{Dong:2017iw,Manzeli:2017ib} and many others. Here, the electronic structure is ranged from metallic (graphene, silicene) to insulating ($h$-BN) state. Although a great quantity of 2D crystals have been widely explored, most of them are lacking of intrinsic ferromagnetic (FM) ordering. Inspired by the discovery of layer-dependent ferromagnetism in insulating CrI$_3$ monolayer with a Curie temperature ($T_\mathrm{C}$) of $45$\,K~\cite{Huang:2017kd}, many 2D magnetic materials have been recently synthesized, such as semiconducting Cr$_2$Ge$_2$Te$_6$~\cite{Gong:2017jf}, its metallic analogue Fe$_3$GeTe$_2$~\cite{Deng:2018cp}, and semiconducting MnSe$_2$~\cite{OHara:2018hi}, initiating enormous attention to the field of magnetic 2D atomic crystals~\cite{Burch:2018gr,Gibertini:2019cp}. Meanwhile, some novel properties have been also predicted for the 2D magnetic materials, e.\,g. the Dirac spin-gapless semiconductor state for NiCl$_3$ and $h$-V$_2$O$_3$ monolayers~\cite{He:2017ib,Gog:2019fl}, topological magnetic-spin textures in Cr$_2$Ge$_2$Te$_6$~\cite{Han:2019gr}, spontaneous valley splitting in 2D VAgP$_2$Se$_6$~\cite{Song:2018ck}, etc. Thus, the rapidly advancing progress in the field of the studies of magnetism in pure 2D materials offers great opportunities for new physical paradigms and next generation information technology~\cite{Burch:2018gr}.
\vspace{-0.2cm}

Currently, increased research attention is focused on transition-metal phosphorus trichalcogenides (MPX$_3$ with M = transition metal and X = S, Se), which crystals can be easily exfoliated into monolayers~\cite{Du:2016ft}, where a single 2D unit consisting of the transition metal atoms shows honeycomb lattice structure similar to that of graphene (Fig.~\ref{structure}a,b). However, unlike graphene, which has a zero band gap, members of the MPX$_3$ family demonstrate wide variation of  band gaps ranging from $0.5$\,eV to $3.5$\,eV~\cite{Wang:2018dh}. In addition, transition metal compounds have a large spin-orbit coupling and strong electronic correlations compared to the case of graphene.

\begin{figure}
\vspace{0.2cm}
\includegraphics[width=\textwidth]{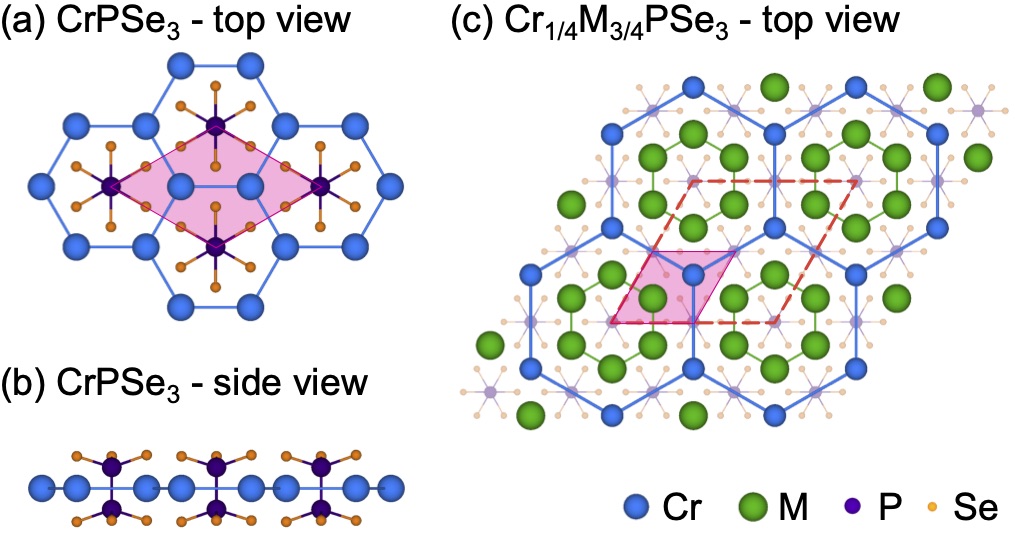}
\caption{(a,b) Top and side views of the crystal structure of the 2D CrPSe$_3$ monolayer.  (c) Top view of Cr$_{1/4}$M$_{3/4}$PSe$_3$ with the metal ions highlighted. Small shaded rhombus indicates the ($1\times1$) unit cell of CrPSe$_3$. Large dashed rhombus indicates the unit cell of mixed  compound, which has a ($2\times2$) periodicity with respect to the pristine CrPSe$_3$. Spheres of different size/color represent ions of different type. }
\label{structure}
\end{figure}

Overall, due to the diversity of fully or partially occupied $d$-orbitals for the transition metal ions, one may expect a variety of magnetic behavior for MPX$_3$ monolayers. In this regard, 2D monolayers were studied mostly by theoretical methods, excepting the case of M = Fe for which experimental results are available as well~\cite{Lee:2016ga}. 
According to the recent theoretical studies, most of 2D MPX$_3$ monolayers have antiferromagnetic (AFM) arrangement of magnetic moments of transition metal ions in their ground states~\cite{Chittari:2016cd,Sivadas:2015gq,Kim:2018gj,Yang:2020ex}.

\begin{figure*}
\includegraphics[width=\textwidth]{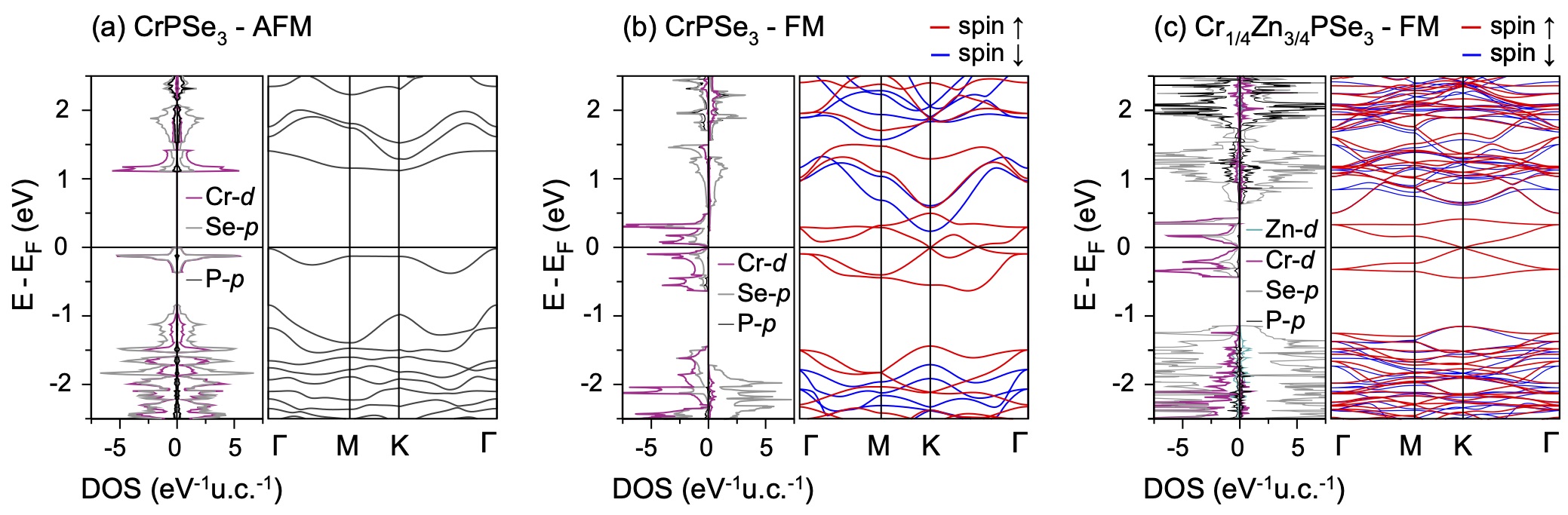}
\caption{Band structures and orbital-projected density of states calculated for (a) 2D CrPSe$_3$ monolayer in the AFM configuration; (b) 2D CrPSe$_3$ monolayer in the FM configuration; (c) 2D Cr$_{1/4}$Zn$_{3/4}$PSe$_3$ monolayer in the FM configuration. }
\label{dos+bands_CrPSe3}
\vspace{-0.2cm}
\end{figure*}

The magnetic phase transition for the MPX$_3$ materials to the FM state can be realized through applying stress~\cite{Chittari:2016cd} or modulating the carrier concentration~\cite{Li:2014de}. For instance, it was theoretically shown that by means of applying an external voltage gate 2D MnPX$_3$ (X = S, Se) can be converted from the AFM semiconducting state into the half-metallic ferromagnetic (HMF) state with $T_\mathrm{C} = 206$\,K~\cite{Li:2014de}. A further approach to alter the magnetic properties is alloying. Here the properties can be tailored with variation of the concentration of an alloying extent~\cite{Masubuchi:2008cx,Goossens:2013ke}.

In this manuscript, the electronic and magnetic properties of 2D CrPSe$_3$ monolayers are studied using density functional theory including an on-site Coulomb term (DFT$+U$) and tight-binding approach (TBA). While CrPSe$_3$ demonstrates the AFM ground state, its alloying with MPSe$_3$ (M = Zn, Cd, Hg) can be used to stabilize the FM state. Cr$_{1\textrm{-}x}$M$_x$PSe$_3$  ($x = 3/4$) is the HMF (Dirac spin-gapless semiconducting) state with near room temperature $T_C$, obtained in the Monte Carlo simulations. This HMF state emerges Dirac cones in the first 2D Brillouin zone, which originate from the long-range exchange interactions between the $d$-orbitals of Cr$^{2+}$ ions formed a honeycomb lattice. If spin-orbit coupling (SOC) is taken into account, the Cr$_{1\textrm{-}x}$M$_x$PSe$_3$ monolayers become an intrinsic Chern insulators with a large non-trivial band gaps. The high Curie temperature and single-spin Dirac states with high Fermi velocities make 2D  Cr$_{1\textrm{-}x}$M$_x$PSe$_3$ monolayers very promising materials for low-dimensional spintronics applications.

\begin{table*}
\caption{Results for the 2D CrPSe$_3$ monolayer and for the 2D Cr$_{1/4}$M$_{3/4}$PSe$_3$ (M = Zn, Cd, Hg) monolayers obtained for the FM state: $\Delta E_\mathrm{tot}=E_\mathrm{FM}-E_\mathrm{AFM}$ (in meV/u.c.) is the energy difference between the FM and AFM states; band gaps, $E_g$ (in meV), are given for the spin-up ($\uparrow$) and spin-down ($\downarrow$) channels; $J_3$ (in meV) is the exchange coupling parameter between two local spins; $M_\mathrm{Cr}$ (in $\mu_B$) is  Cr magnetic moment; T (in K) is a critical temperature; MAE (in meV/u.c.) is magnetic anisotropy energy; $v_F$ (in m/s) is Fermi velocity. }
\label{tab1}
\begin{ruledtabular}
\begin{tabular}{l l l r c c c c }
System	& $\Delta E_\mathrm{tot}$  &$E_g$ & $J_3$  &$M_\mathrm{Cr}$  &$T$ & MAE  & $v_F$ \\[0.1cm]
\hline\\[-0.2cm]
CrPSe$_3$					&$285\qquad$	&$\uparrow\,\,0/0^a$ ($42.8/14.7$)$^b$; $\downarrow\,\,2079$	&$0.39$ 	&$3.81$	&--	&$650.3$	&$16.11/7.98^a \times10^5$		\\[0.2cm]
Cr$_{1/4}$Zn$_{3/4}$PSe$_3$ 	&$-451\qquad$	&$\uparrow\,\,0$ ($8.6$)$^b$; $\downarrow\,\,1650$		&$5.03$ 	& $3.87$ 	&$264$	& $37.0$	&  $3.87\times10^5$	\\[0.2cm]
Cr$_{1/4}$Cd$_{3/4}$PSe$_3$ 	&$-416\qquad$	&$\uparrow\,\,$ $0$ ($8.2$)$^b$; $\downarrow\,\,1660$	&$4.39$ 	& $3.98$ 	&$235$	& $119.7$	&  $3.98\times10^5$	\\[0.2cm]
Cr$_{1/4}$Hg$_{3/4}$PSe$_3$ 	&$-400\qquad$	&$\uparrow\,\,0$ ($6.3$)$^b$; $\downarrow\,\,970$		&$4.18$ 	& $3.99$ 	&$220$	& $87.2$	&  $4.03\times10^5$	\\[0.2cm]
\end{tabular}
\end{ruledtabular}
\raggedright
\footnotesize{
$^a$at K and K/2, respectively; $^b$for the data in parenthesis SOC is taken into account}
\end{table*}
 
Our DFT$+U$ calculations (see Supplemental Material~\cite{Suppl:XX} for computational details) show that 2D CrPSe$_3$ crystal prefers the AFM coupling in its ground state with $T_N=105$\,K (see Tab.~S1, Fig.~S1, and Fig.~S2 in Supplemental Material~\cite{Suppl:XX}). The high-spin configuration of Cr$^{2+}$ results in ${M}_\mathrm{Cr}=3.8\,\mu_B$. The calculated electronic band structure indicates that the 2D CrPSe$_3$ crystal is a semiconductor with an indirect band gap of $1.11$\,eV (Fig.~\ref{dos+bands_CrPSe3}a). Due to the quantum confinement effects this value is slightly larger than the recently published band gap value of a bulk phase ($0.70$\,eV)~\cite{Dedkov:2020ca}. The bands in the vicinity of Fermi energy are mostly composed of Cr-$d$ and Se-$p$ orbitals. Hence, the $p$--$d$ exchange interactions are much weaker than the direct $d$--$d$ exchange interactions between the NN Cr ions with a bond $d_\mathrm{Cr-Cr}=3.667$\,\AA, leading to the high negative value of $J_1 = -2.03$\,meV (see Supplementary Material~\cite{Suppl:XX} for additional theoretical details). For the 2NN and 3NN Cr-ions, the direct $d$--$d$ interactions decrease significantly with increasing distance between Cr-ions and considerable long-range $p$--$d$ super-exchange interactions result in positive values of $J_2 = 0.12$\,meV and $J_3 = 0.39$\,meV. Thus, the origin of the AFM ground-state can be attributed to the competition between the NN AFM direct Cr--Cr ($d$--$d$) exchange interactions and the indirect Cr--Se$\cdots$Se--Cr ($p$--$d$) superexchange interactions. 

The FM state of  2D CrPSe$_3$ is $285$\,meV higher in energy than the AFM ground-state (Tab.~\ref{tab1}). Still, it attracts our attention due to its exciting band structure, which demonstrates a half-metallic state with the metallic spin-up channel and insulating spin-down channel (Fig.~\ref{dos+bands_CrPSe3}b). For the spin-up electrons, the linear dispersion is observed in the vicinity of the Fermi level and multiple cones appear: one at the K point and an extra one around the midpoint of the $\Gamma$ -- K line named as K/2. The spin-up states near the Fermi level are dominated by Cr-$d_{xz}$ and Cr-$d_{yz}$ orbitals (Fig.~\ref{TBA_CrPSe3}a), which are hybridized with the $p$ orbitals of Se. The calculations with SOC included lead to the gap opening of $42.8$\,meV and $14.7$\,meV for spin-up states at the K and K/2 points, respectively  (see Supplemental Material~\cite{Suppl:XX} Fig.~S3a for the calculated band structures). 

In order to better understand the observed phenomena a TBA was applied to the studied system (see Supplementary Material~\cite{Suppl:XX} for additional theoretical details). Here the $d_{xz}$ and $d_{yz}$ orbitals can be used as a basis for the honeycomb lattice formed by Cr$^{2+}$ ions and the hopping parameters $t_{1j}$ (NN), $t_{2j}$ (2NN), and $t_{3j}$ (3NN) are marked in Fig.~\ref{TBA_CrPSe3}e for this lattice. The modelled band structures are presented in Fig.~\ref{TBA_CrPSe3}c and they are in reasonable agreement with DFT-calculated band structure (Fig.~\ref{TBA_CrPSe3}a). Interestingly, the Dirac cones are accurately located at the K point and K/2 midpoint with vertexes just crossed by the Fermi level when considering only the 3NN hopping elements. From a comparison between the two different TBA band structures, the $t_{1j}$ and $t_{2j}$ interactions were found to introduce hole pockets at the K/2 point and electron pockets at the K point, respectively. Since the lattice sites connected by 3NN bonds also form a honeycomb structure with a twice larger lattice constant, the Dirac cones at the K points are folded to the midpoints along the $\Gamma$ -- K path. The TBA results confirm that the robust Dirac cones are strongly protected by the lattice symmetry, especially, by the mirror symmetry along the $\Gamma$ -- K path. Thus, the origin of the observed Dirac cones can be attributed to the honeycomb lattice structure composed from chromium ions.

\begin{figure*}
\includegraphics[width=\textwidth]{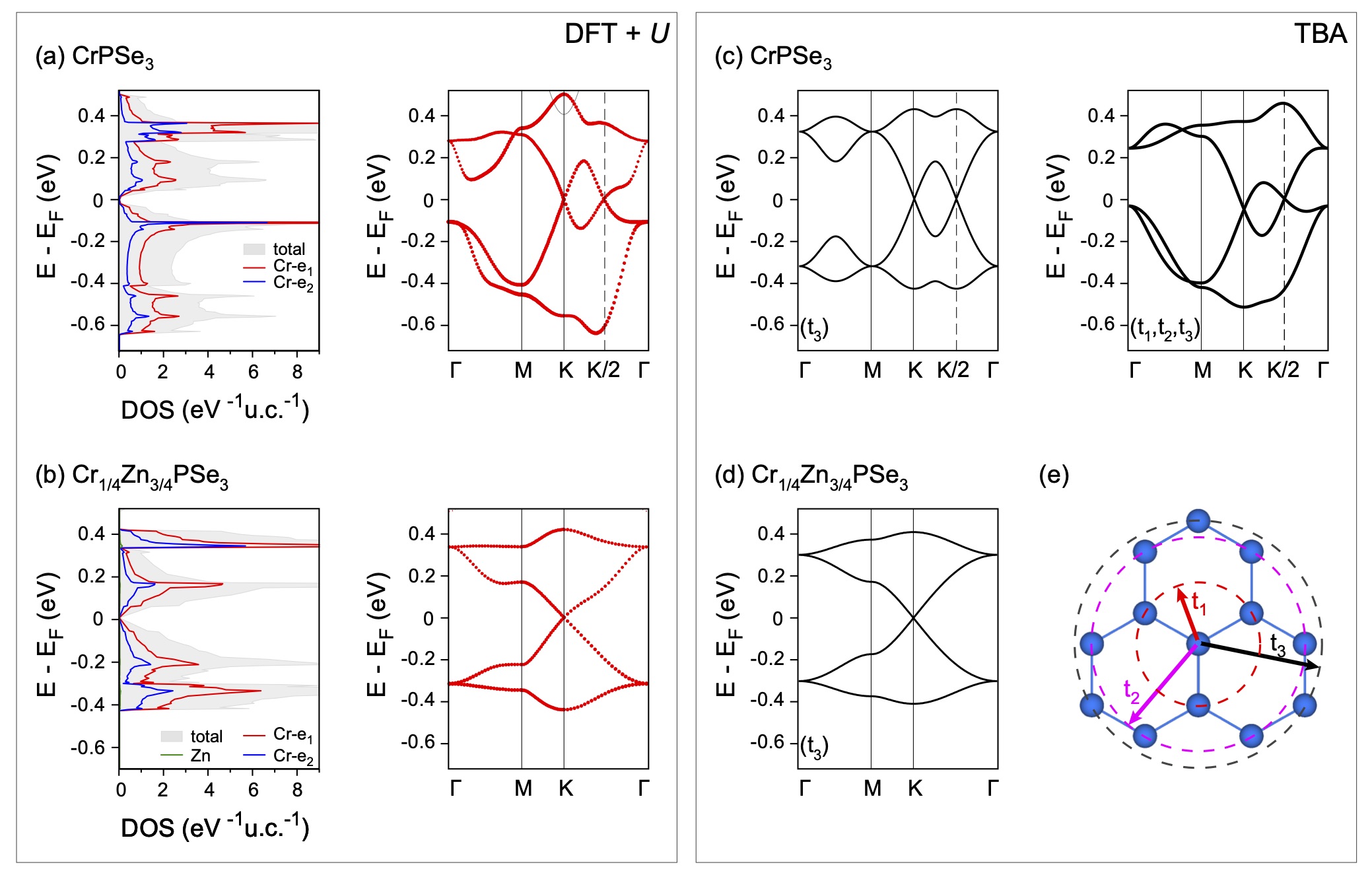}
\caption{(a,b) DFT$+U$ results:
(a) Density of states (left) and band structure (right) of the 2D CrPSe$_3$ monolayer (FM state) in the vicinity of $E_F$.   (b) Density of states (left) and band structure (right) of the 2D Cr$_{1/4}$Zn$_{3/4}$PSe$_3$ monolayer in the vicinity of $E_F$. Grey shadowed area in DOS shows the total DOS; red and blue lines show the Cr $e_1$ ($d_{xz}$, $d_{yz}$) and Cr $e_2$ ($d_{x^2-y^2}$, $d_{z^2}$, $d_{xy}$) PDOS, respectively; green line corresponds to the Zn $d$ PDOS. In the band structure the weight of the Cr $e_1$ ($d_{xz}$, $d_{yz}$) states  is proportional to the width of the colored line.
(c-e) TBA results:
(c) Bands for 2D CrPSe$_3$ monolayer when considering either only $t_3$ hopping elements (left) or $t_1$, $t_2$, $t_3$ hopping elements (right). (d) Bands for 2D Cr$_{1/4}$Zn$_{3/4}$PSe$_3$ monolayer.  (e) Hopping elements $t_1$, $t_2$, $t_3$.  }
\label{TBA_CrPSe3}
\vspace{-0.2cm}
\end{figure*}

In the next step we intend to find a way for stabilisation of the FM state in CrPSe$_3$. Here we propose to use its alloying with MPSe$_3$ with M = group-12 metals. The 2D ZnPSe$_3$, CdPSe$_3$, and HgPSe$_3$ are paramagnetic compounds, which correlates with the $d^{10}s^2$-configuration of Zn, Cd, and Hg. They are semiconductors with the calculated indirect gaps of $1.47$\,eV, $1.49$\,eV, $0.83$\,eV, respectively (see Tab.~S2 in Supplemental Material~\cite{Suppl:XX}). These DFT$+U$ values are underestimated with respect to the available experimental data ($E_g^\mathrm{CdPSe_3} = 2.29$\,eV~\cite{Calareso:1997jk}, $E_g^\mathrm{HgPSe_3} = 1.95$\,eV~\cite{Calareso:1997fi}), but in agreement with the previously published theoretical results~\cite{Xiang:2016gu}. The band edges of these monolayers are composed from Se-$p$ orbitals, whereas the completely filled $d$-states are strongly localized at around $8$\,eV below the Fermi level (see Fig.~S4 in Supplemental Material~\cite{Suppl:XX}). 

As shown above, the origin of the AFM ground-state of 2D CrPSe$_3$ can be attributed to the competition between the NN AFM direct Cr--Cr ($d$--$d$) exchange interactions and the indirect Cr--Se$\cdots$Se--Cr ($p$--$d$) superexchange interactions. Consequently, FM may arise when the direct $d$--$d$ coupling is significantly weakened or even entirely absent. This can be achieved if all nearest-neighbour Cr ions will be substituted by group-12 metal ions, i.\,e. by means of alloying of CrPSe$_3$ and MPX$_3$ with the ratio \mbox{Cr : M = 1 : 3}. The proposed structure for such an alloy is presented in Fig.~\ref{structure}c. 

In order to check our idea, we performed the electronic structure calculations for the suggested alloy, i.\,e. Cr$_{1/4}$M$_{3/4}$PSe$_3$. As expected, all systems under study prefer the FM ground states (Tab.~\ref{tab1}). In order to further confirm our idea, we carried out similar calculations for the opposite case,  i.\,e. Cr$_{3/4}$M$_{1/4}$PSe$_3$, and observed stabilisation of the AFM coupling (see Fig.~S5 and Tab.~S4 in Supplemental Material~\cite{Suppl:XX}). The spin-resolved band structure of Cr$_{1/4}$Zn$_{3/4}$PSe$_3$ in the ground-state configurations is presented in Fig.~\ref{dos+bands_CrPSe3}c.  The band structures for the cases of M = Cd and Hg show similar behavior and can be seen in Supplemental Material~\cite{Suppl:XX} (Fig.~S5). All systems under consideration show the HFM behavior with the metallic spin-up channel and insulating spin-down one. The Dirac fermions emerge at the high-symmetry K point in the vicinity of the Fermi level. These Dirac cones are spin-polarized and resulted from the strong hybridization between Cr-$d$ and Se-$p$ orbitals. The Cr--Cr interactions are still strong enough in these systems, despite of a long length larger than $7.00$\,\AA\ for the Cr-honeycomb lattices. Still the FM $p$ -- $d$ superexchange interactions prevail resulting in large $J_3$ (3NN Cr--Cr pair) values of $5.03$\,meV, $4.39$\,meV, and $4.18$\,meV, obtained for Cr$_{1/4}$M$_{3/4}$PSe$_3$ with M = Zn, Cd, and Hg, respectively (Tab.~\ref{tab1}). Using the calculated exchange parameters we have employed Monte-Carlo simulations based on the Ising model and estimated Curie temperatures ($T_\mathrm{C}$) (see Supplementary Material~\cite{Suppl:XX}, Fig.~S6). The calculated $T_\mathrm{C}$ listed in Tab.~\ref{tab1} suggest the possibility of using the studied monolayers in near-room temperature spintronics.

The electronic structure of Cr$_{1/4}$Zn$_{3/4}$PSe$_3$ shows rather rare Dirac spin-gapless semiconductor characteristics that are essential for potential high-speed spin filter devices. The calculated Fermi velocities ($v_F$) in the $\Gamma$ -- K direction for Dirac electrons at the Fermi level are: $3.87\times10^5\,\mathrm{m\,s}^{-1}$, $3.98\times10^5\,\mathrm{m\,s}^{-1}$, and $4.03\times10^5\,\mathrm{m\,s}^{-1}$ for Cr$_{1/4}$M$_{3/4}$PSe$_3$ with M = Zn, Cd, and Hg, respectively. These values for $v_F$ are about a half of the graphene value of $8.2\times10^5\,\mathrm{m\,s}^{-1}$~\cite{Malko:2012jn}, demonstrating the excellent electronic transport properties of Cr$_{1/4}$M$_{3/4}$PSe$_3$ monolayers. Since the Dirac electrons are derived from the strong $p$ -- $d$ hybridization, the SOC effect was considered to open a band gap. The SOC-induced estimated gaps are $8.6$\,meV, $8.2$\,meV and $6.3$\,meV for Cr$_{1/4}$M$_{3/4}$PSe$_3$ with M = Zn, Cd, and Hg, respectively (see Supplementary Material~\cite{Suppl:XX}, Fig.~S3b-d). These gaps are significantly larger than that of the other Dirac materials, such as graphene~\cite{Gmitra:2009fh,Abdelouahed:2010gq}, which are characterized by Dirac states composed of $p$-orbitals with weak spin-orbital couplings.

As for pristine CrPSe$_3$, TBA with $d_{xz}$ and $d_{yz}$ orbitals used as a basis for the honeycomb lattice formed by Cr$^{2+}$ ions was applied for the systems under study (see Tab.~S3 in Supplementary Material~\cite{Suppl:XX}). The band structure obtained for  M = Zn is presented in Fig.~\ref{TBA_CrPSe3}d and it is in a good agreement with DFT-calculated band structure (Fig.~\ref{TBA_CrPSe3}b). Especially the Dirac cone at the K point as well as the bandwidth calculated with TBA are well reproduced. The hopping parameter $t_{11}$ controls the band width while $t_{12}$ affects the band dispersion, respectively. Some differences between TBA and DFT band structures are caused by the neglecting of the contributions from Se $p$ orbitals.

\textit{In conclusion}, the electronic structure of 2D CrPSe$_3$ monolayer was explored using first-principles calculations. The calculated band structures obtained on the DFT$+U$ level illustrate that the 2D CrPX3 monolayers adopt the AFM ground state. Simultaneously, its FM state is half-metallic ferromagnet with multiple spin-polarized Dirac cones. Deep analysis of the observed  phenomenon allowed us to find the way to stabilize the FM state by means of alloying of the studied system with MPSe$_3$, where M is group-12 metal (Zn, Cd, Hg). Depending on the concentration and location of the alloying extend, it is possible to achieve the HMF Dirac spin-gapless semiconducting state in the FM phase for Cr$_{1\textrm{-}x}$M$_x$PSe$_3$  ($x = 3/4$) monolayers. For such cases, the Dirac cone is located at the K point of the Brillouin zone and these fully spin-polarized Dirac cones originate from the long-range exchange interactions between the $d$-orbitals of Cr$^{2+}$ ions formed a honeycomb lattice. Consequently, SOC opens sizable energy gaps for fully spin-polarized Dirac cones, which are significantly larger compared to the case of Dirac states composed of $p$-orbitals. The estimated rather high Curie temperatures, large SOC gaps, high Fermi velocities, and single-spin Dirac states for Cr$_{1\textrm{-}x}$M$_x$PSe$_3$  ($x = 3/4$) monolayers give rise to great expectations for its potential applications in spintronics.

\medskip
This work was supported by the National Natural Science Foundation of China (Grant No. 21973059). J.Y. was supported by the Natural Science Foundation of Hubei Province (Grant No. 2018CFB724) and of Education Department (Grant No. D20171803).

\clearpage

\noindent
\textbf{Supplementary material for the manuscript: Dirac fermions in half-metallic ferromagnetic mixed Cr$_{1\textrm{-}x}$M$_x$PSe$_3$ monolayers}

\author{Juntao Yang,$^{1,2}$ Yong Zhou,$^{1}$ Yuriy Dedkov,$^{1,}$\footnote{E-mail: dedkov@shu.edu.cn} and Elena Voloshina$^{1,}$\footnote{E-mail: voloshina@shu.edu.cn}}

\affiliation{$^1$Department of Physics, Shanghai University, Shangda Road 99, 200444 Shanghai, China}
\affiliation{$^2$School of Science, Hubei University of Automotive Technology, 167 Checheng West Road, Shiyan City, 442002 Hubei, China}

\date{\today}

\maketitle

\noindent \textbf{Content:}
\begin{itemize}

\item[1.]  Computational details.

\item[2.]  Calculation of exchange-coupling parameters: $J_1$, $J_2$, $J_3$.

\item[3.] Monte-Carlo simulations.

\item[4.] TBA model.

\item[5.] Tab.~S1: Total energies ($E_\mathrm{tot}$, in eV per ($2\times2$) unit cell) as well as total energies relative to that of the lowest-energy magnetic configuration ($\Delta E$, in meV per unit cell) calculated using the PBE$+U$ ($U=4$\,eV) approximation for the ferromagnetic (FM), N\'eel antiferromagnetic (AFM), zigzag antiferromagnetic (zAFM), and stripy antiferromagnetic (sAFM) configurations of single-layer CrPSe$_3$.

\item[6.] Tab.~S2: Results for the pristine 2D MPSe$_3$ monolayers: Ground state magnetic configuration; optimized lattice parameters: in-plane lattice parameter ($a$), layer thickness ($h$), M--M and M--Se distances ($d_\mathrm{M-M}$ and $d_\mathrm{M-Se}$,  band gap ($E_g$).

\item[7.] Tab.~S3: The hopping parameters, onsite energies and band widths for Cr$_{1/4}$M$_{3/4}$PSe$_3$ (M = Zn, Cd, Hg).

\item[8.] Tab.~S4: Results for the 2D Cr$_{1-x}$M$_x$PSe$_3$ (M = Zn, Cd, Hg, $x = 1/4, 3/4$) monolayers obtained for their ground-state states: $\Delta E_\mathrm{tot} = E_\mathrm{FM} - E_\mathrm{AFM}$ (in meV/u.c.) is the energy difference between the FM and AFM states; band gaps, $E_g$ (in eV), are given for the spin-up ($\uparrow$) and spin-down ($\downarrow$) channels; $J_\mathrm{Cr-Cr}$ (in meV) is the exchange coupling parameter between two local spins; $M_\mathrm{Cr}$ (in $\mu_B$) is Cr magnetic moment; $T$ (in K) is a critical temperature.

\item[9.] Fig.~S1: Four different magnetic configurations of 2D CrPSe$_3$: (a) ferromagnetic (FM), (b) N\'eel antiferromagnetic (AFM), (c) zigzag antiferromagnetic (zAFM), and (d) stripy antiferromagnetic (sAFM).

\item[10.] Fig.~S2: The Monte-Carlo simulated specific heat capacity as a function of temperature for 2D CrPSe$_3$.

\item[11.] Fig.~S3: Comparison of band structures of FM-CrPSe$_3$ as well as mixed Cr$_{1/4}$M$_{3/4}$PSe$_3$ monolayers in the vicinity of $E_F$ obtained with DFT$+U$ and with DFT$+U$+SOC. 

\item[12.] Fig.~S4: Band structures of 2D MPSe$_3$ (M = Zn, Cd, Hg).

\item[13.] Fig.~S5: Band structures and orbital-projected density of states calculated for 2D Cr$_{1-x}$M$_{x}$PSe$_3$ (M = Zn, Cd, Hg; $x = 1/4, 3/4$) monolayers in their ground-state configurations.

\item[14.] Fig.~S6: The Monte-Carlo simulated specific heat capacity as a function of temperature for 2D Cr$_{1/4}$M$_{3/4}$PSe$_3$ (M = Zn, Cd, Hg). 

\end{itemize}

\linespread{1.2}

\clearpage

\section*{1. Computational details}

Spin-polarised DFT calculations based on plane-wave basis sets of $500$\,eV cutoff energy were performed with the Vienna \textit{ab initio} simulation package (VASP)~\cite{Kresse:1996kg,Kresse:1994cp,Kresse:1993hw}. The Perdew-Burke-Ernzerhof (PBE) exchange-correlation functional~\cite{Perdew:1997ky} was employed. The electron-ion interaction was described within the projector augmented wave (PAW) method~\cite{Blochl:1994fq} with Cr ($3p$, $3d$, $4s$), P ($3s$, $3p$), and Se ($4s$, $4p$) states treated as valence states. The Brillouin-zone integration was performed on $\Gamma$-centred symmetry reduced Monkhorst-Pack meshes using a Gaussian smearing with $\sigma = 0.05$\,eV, except for the calculation of total energies. For these calculations, the tetrahedron method with Bl\"ochl corrections~\cite{Blochl:1994ip} was employed. A $12\times 12\times 1$ $k$-mesh was used in the case of ionic relaxations and $24\times 24\times 1$ for single point calculations, respectively. The DFT+$\,U$ scheme~\cite{Anisimov:1997ep,Dudarev:1998dq} was adopted for the treatment of Cr $3d$ orbitals, with the parameter $U_\mathrm{eff}=U-J$ equal to $4$\,eV. Dispersion interactions were considered adding a $1/r^6$ atom-atom term as parameterised by Grimme (``D2'' parameterisation)~\cite{Grimme:2006fc}. To ensure decoupling between periodically repeated layers, a vacuum space of $20$\,\AA\ was used. During structure optimisation, the convergence criteria for energy and force were set equal to $10^{-6}$\,eV and $1\times10^{-2}$\,eV/\AA, respectively.  

\clearpage

\section*{2. Calculation of exchange-coupling parameters: $J_1$, $J_2$, $J_3$}

\begin{figure}
\includegraphics[width=\textwidth]{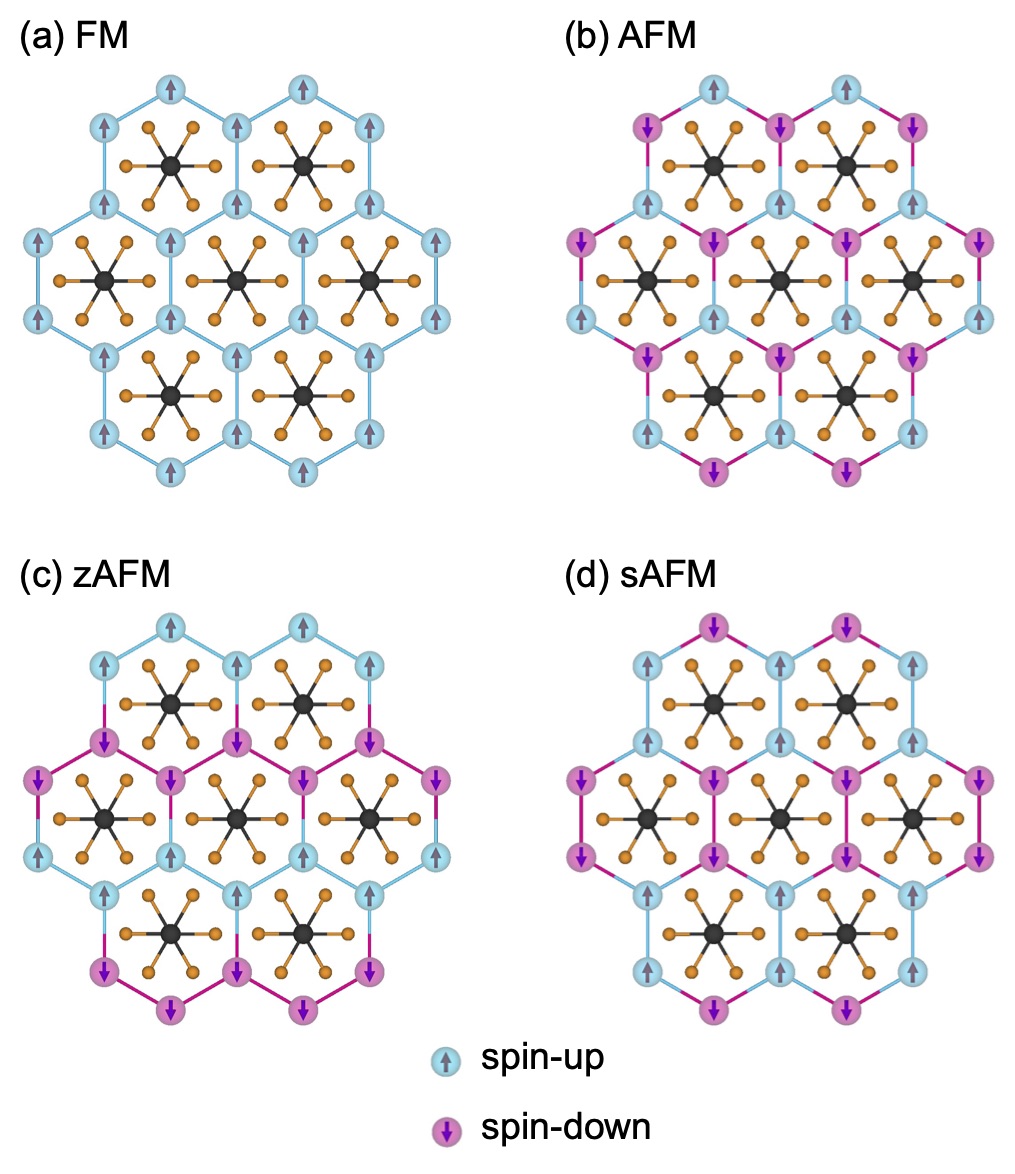}
\noindent
Fig.\,S1. Four different magnetic configurations of 2D CrPSe$_3$: (a) ferromagnetic (FM), (b) N\'eel antiferromagnetic (AFM), (c) zigzag antiferromagnetic (zAFM), and (d) stripy antiferromagnetic (sAFM).
\label{mconfig}
\end{figure}

The magnetic coupling parameter $J$ can be extracted by mapping the total energies of four spin orders (Fig.~\ref{mconfig}) to the Ising model with a Hamiltonian
\begin{equation}\nonumber
H=\sum_{\langle i,j \rangle}J_1\vec{S}_i \cdot \vec{S}_j+\sum_{\langle \langle i,j \rangle \rangle}J_2\vec{S}_i \cdot \vec{S}_j+\sum_{\langle \langle \langle i,j \rangle \rangle \rangle}J_3\vec{S}_i \cdot \vec{S}_j,
\end{equation}
where $\vec{S}_i$ is the net spin magnetic moment of the Cr ions at site $i$, three different distance magnetic coupling parameters were estimated, considering one central Cr ions interacted with three nearest neighbouring (NN, $J_1$), six next-nearest neighbouring (2NN, $J_2$), and three third-nearest neighbouring (3NN, $J_3$) Cr ions, respectively. 

 The long-range magnetic exchange parameters ($J$) can be obtained as \cite{Sivadas:2015gq}
\begin{equation}\nonumber
\begin{aligned}
 &J_1=\frac{E_\mathrm{FM}-E_\mathrm{AFM}+E_\mathrm{zAFM}-E_\mathrm{sAFM}}{8S^2}\,,\\
 &J_2=\frac{E_\mathrm{FM}+E_\mathrm{AFM}-(E_\mathrm{zAFM}+E_\mathrm{sAFM})}{16S^2}\,, \\
 &J_3=\frac{E_\mathrm{FM}-E_\mathrm{AFM}-3(E_\mathrm{zAFM}-E_\mathrm{sAFM})}{24S^2}\,.
 \end{aligned}
 \end{equation}
 
 For the mixed compound, considering only one possible magnetic interaction, the exchange parameters are calculated as $J=\Delta E/6S^2$, where $\Delta E$ is the energy difference between the FM and AFM orders: $\Delta E = E_\mathrm{FM}-E_\mathrm{AFM}$.

\clearpage

\section*{3. Monte-Carlo simulations}

To estimate $T_\mathrm{N}$ and $T_\mathrm{C}$ temperatures, Monte-Carlo (MC) simulations were performed within the Metropolis algorithm with periodic boundary conditions~\cite{Metropolis:1953in}. 

Using the exchange parameters, MC simulations based on the Ising model were then carried out to evaluate the N\'eel temperatures in a $32 \times 32$ 2D superlattice using at least $10^9$ steps. In each step, the spins on all the sites in the superlattice flip randomly according to the spin states. The specific heat capacity $C_v(T) = (\langle E^2 \rangle - \langle E \rangle^2)/k_BT^2$ as a function of temperature were calculated. The critical temperatures were extracted from the peak position of the specific heat capacity $C_v(T)$.

\clearpage

\section*{4. TBA model}

The effective Hamiltonian can be denoted as~\cite{Gu:2019kw}
\begin{equation}\nonumber
H_0=\sum_{k}\psi_{\kappa}^{\dag}h_{\kappa}\psi_{\kappa}=\sum_{\kappa\alpha\beta}\sum_{\mu\nu}h_{\mu\nu}^{\alpha\beta}(\kappa)c^{\dag}_{\alpha\mu}c_{\beta\nu}\,.
\end{equation} 
Here $\mu$, $\nu$ indicate the orbitals ($d_{xy}$, $d_{yz}$), and $\alpha$, $\beta$ indicate two sublattices with different orientations ($A$, $B$). The basic $\psi_{\kappa}^{\dag}=(c^{\dag}_{\kappa,A,xz},c^{\dag}_{\kappa,A,yz},c^{\dag}_{\kappa,B,xz},c^{\dag}_{\kappa,B,yz})$ with the electron creation operator $c^{\dag}_{\alpha\mu}$ means creating an electron for $\mu$ orbital with momentum $\mathbf{k}$ in $\alpha$ sublattice. The $4\times4$ basis $h(\kappa)$ is described as
\begin{equation}\nonumber
h(k)=\begin{pmatrix}
\omega-\epsilon & \gamma \\ 
 \gamma^{\dag} & \omega^{T}-\epsilon 
\end{pmatrix}
\end{equation}
where $\epsilon$ is onsite energy, $T=C_jT^iC_j^{-1}$ is a $2\times 2$ matrix of the hopping parameter $t_{ij}$ with a three-fold rotation operation $C_j$ to the $ij$ hopping direction between the manifold of $d$ orbitals. The element $\omega_k=\sum_{{\langle \langle i,j \rangle \rangle}}{e^{-i\mathbf{k\cdot\,a_{2j}}}T^{2NN}_j}$ is the summation of Fourier transform of t$_{2j}$ hopping matrix with 2NN neighboring vector $\mathbf{a_2}$, similarly $\gamma$ refers to the summations of t$_{1j}$ and 3NN t$_{3j}$ hopping matrices Fourier transformation. 

The matrix elements of the $4\times 4$ Hamiltonian are
\begin{align*}
h_{11}=&2t_{21}\cos^2\frac{k_x}{2}+(t_{21}+3t_{23})\cos\frac{k_x}{2}\cos\frac{\sqrt{3}k_y}{2}-2t_{21}\sin^2\frac{k_x}{2}\,,\\[0.5cm]
h_{12}=&\sin\frac{k_x}{2}[4it_{22}\cos\frac{k_x}{2}-4it_{22}\cos\frac{\sqrt{3}k_y}{2}+\sqrt{3}(-t_{21}+t_{23})\sin\frac{\sqrt{3}k_y}{2}]\,,\\[0.5cm]
h_{13}=&\frac{1}{2}(\cos\frac{k_y}{2\sqrt{3}}-i\sin\frac{k_y}{2\sqrt{3}})(t_{11} +3 t_{12}) \cos\frac{k_x}{2} \\
&+\frac{1}{2}(\cos\frac{k_y}{2\sqrt{3}}-i \sin\frac{k_y}{2\sqrt{3}})[i(2(t_{11}-t_{31}) + (t_{31} + 3 t_{32}) \cos k_x]\sin\frac{\sqrt{3}k_y}{2}\\
&+\frac{1}{2}(\cos\frac{k_y}{2\sqrt{3}}-i \sin\frac{k_y}{2\sqrt{3}}) [2 (t_{11} + t_{31}) + (t_{31} + 3 t_{32}) \cos k_x] \cos\frac{\sqrt{3}k_y}{2}\,,\\[0.5cm]
h_{14}=&\frac{\sqrt{3}}{2}[i(t_{11} - t_{12}) \cos\frac{k_y}{2\sqrt{3}} \sin\frac{k_x}{2}-i(t_{31}-t_{32}) \cos^2\frac{k_y}{2\sqrt{3}} \sin k_x\\
& +(t_{11} - t_{12}) \sin\frac{k_x}{2} \sin\frac{k_y}{2\sqrt{3}}-\frac{1}{2}i(t_{31} - t_{32}) (\cos\frac{k_y}{\sqrt{3}}\sin k_x + 2 i \sin\frac{k_y }{\sqrt{3}}\sin k_x-\sin k_x)]\,,\\
\end{align*}
\begin{align*}
h_{22}=&2t_{23}\cos^2\frac{k_x}{2} + (t_{23} + 3t_{21}) \cos\frac{k_x}{2}\cos\frac{\sqrt{3}k_y}{2}-2t_{23}\sin^2\frac{k_x}{2}\,,\\[0.5cm]
h_{24}=&\frac{1}{2} (\cos\frac{k_y}{2\sqrt{3}} - i \sin\frac{k_y}{2\sqrt{3}}) [(3 t_{11} + t_{12}) \cos\frac{k_x}{2} + (2 (t_{12} + t_{32}) + (3 t_{31} + t_{32}) \cos k_x) \cos\frac{\sqrt{3}k_y}{2}]\\
&+\frac{1}{2} (\cos\frac{k_y}{2\sqrt{3}} - i \sin\frac{k_y}{2\sqrt{3}})[ 2i (t_{12} - t_{32}) +i(3 t_{31} + t_{32}) \cos k_x\sin\frac{\sqrt{3}k_y}{2}]
\end{align*}

Here, $h_{33}=h_{11}$, $h_{44}=h_{22}$, $h_{23}=h_{14}$ and $h_{34}=h^*_{12}$.

The hopping parameters for TBA were evaluated by Wannier90 program~\cite{Pizzi:2019be}.

The three-different distance hopping parameters of 2D CrPSe$_3$ are $t_{11} =-0.066969$, $t_{12} =-0.048230$, $t_{21} =0.016447$, $t_{22} =-0.002230$, $t_{23}=0.0020415$, $t_{31}=-0.034616$, and $t_{32}=0.241613$, respectively, and the onsite energies are $\epsilon_{xzxz} = -2.275418$\,eV and $\epsilon_{yzyz} = -2.274829$\,eV. Considering only the NN interactions ($t_{2j} = t_{3j} = 0$), the hopping parameters and chemical potentials of M$_{1/4}$Zn$_{3/4}$PSe$_3$ are listed in Tab.~S3, respectively.

\clearpage

\begin{table}
\caption{Total energies ($E_\mathrm{tot}$ in eV per ($2\times2$) unit cell) as well as total energies relative to that of the lowest-energy magnetic configuration ($\Delta E$, in meV per unit cell) calculated using the DFT$+U$ ($U=4$\,eV) approximation for the ferromagnetic (FM), N\'eel antiferromagnetic (AFM), zigzag antiferromagnetic (zAFM), and stripy antiferromagnetic (sAFM) configurations of single-layer CrPSe$_3$. }

\begin{ruledtabular}
\begin{tabular}{lllll }
 Energy     & FM        & AFM	& zAFM	&  sAFM \\
\hline
$E_\mathrm{tot}$	&$-204.016$	&$-204.301$	&$-204.040$	&$-204.224$\\
$\Delta E$			&$285$		&$0$		& $261$         	&$77$\\
\end{tabular}
\end{ruledtabular}

\end{table}

\clearpage

\begin{table*}
\caption{Results for the pristine 2D MPSe$_3$ monolayers: Ground state magnetic configuration; optimized lattice parameters: in-plane lattice parameter ($a$), layer thickness ($h$), M--M and M--Se distances ($d_\mathrm{M-M}$ and $d_\mathrm{M-Se}$),  band gap ($E_g$). }
\begin{ruledtabular}
\begin{tabular}{ll c c c c c}
System		&State	&$a$ (\AA)	&$h$	 (\AA)	&$d_\mathrm{M-M}$	 (\AA)	&$d_\mathrm{M-Se}$ (\AA) 	&$E_g$ (eV)	 \\
\hline
CrPSe$_3$	&AFM	&$6.352$		&$3.534$		&$3.667$					&$2.727$					&$1.11$	\\
ZnPSe$_3$	&NM		&$6.291$ ($6.290$~\cite{Jorgens:2004gv})		&$3.374$		&$3.662$					&$2.839$					&$1.47$	\\ 
CdPSe$_3$	&NM		&$6.515$ 		&$3.580$		&$3.761$ 					&$2.839$					&$1.49$	\\
HgPSe$_3$	&NM	 	&$6.550$ 	($6.545$~\cite{Jandali:1978dp})	&$3.633$ 		&$3.782$ 					&$2.873$ 					&$0.83$	\\	
\end{tabular}
\end{ruledtabular}
\end{table*}

\clearpage

\begin{table*}[h]
	\caption{The hopping parameters, onsite energies and band widths for Cr$_{1/4}$M$_{3/4}$PSe$_3$ (M = Zn, Cd, Hg).}	
\begin{ruledtabular}
\begin{tabular}{lccccccccc}
System &\multicolumn{2}{c}{Hopping parameter (eV) }& \multicolumn{2}{c}{$\epsilon$ (eV)} &\multicolumn{2}{c}{Band width (eV)}  \\	
		& $t_{11}$   & $t_{22}$  & $h_{xzxz}$ & $h_{yzyz}$   & TBA & DFT   \\  
\hline
		
Cr$_{1/4}$Zn$_{3/4}$PSe$_3$   	& 0.236657   & -0.036112  & -2.299083 & -2.298403  & 0.82  & 0.87  \\
		
Cr$_{1/4}$Cd$_{3/4}$PSe$_3$    	& 0.213962   & -0.033283  &-2.757786  & -2.757788  & 0.74  & 0.77   \\ 
		
Cr$_{1/4}$Hg$_{3/4}$PSe$_3$    	& 0.209113  & -0.028072   & -2.815065 & -2.814500  & 0.71  & 0.73   \\  
	\end{tabular}
	\label{TBMHopping}
\end{ruledtabular}
\end{table*}

\clearpage

\begin{table*}
\caption{Results for the 2D Cr$_{1-x}$M$_x$PSe$_3$ (M = Zn, Cd, Hg, $x = 1/4, 3/4$) monolayers obtained for their ground-state states: $\Delta E_\mathrm{tot} = E_\mathrm{FM} - E_\mathrm{AFM}$ (in meV/u.c.) is the energy difference between the FM and AFM states; band gaps, $E_g$ (in eV), are given for the spin-up ($\uparrow$) and spin-down ($\downarrow$) channels; $J_\mathrm{Cr-Cr}$ (in meV) is the exchange coupling parameter between two local spins; $M_\mathrm{Cr}$ (in $\mu_B$) is Cr magnetic moment; $T$ (in K) is a critical temperature.}
\begin{ruledtabular}
\begin{tabular}{l l  r  l r c l }
M & x  &  $\Delta E_\mathrm{tot}$  &$E_g$ & $J_\mathrm{Cr-Cr}$  &$M_\mathrm{Cr}$  &$T$  \\
\hline
Zn	&3/4	&$-451$	&$\uparrow\,\,0$ $\downarrow\,\,1.65$	&$5.03$	&$3.87$	&$264$ \\
	&1/4	&$346$	&$0.78$						&$-4.22$	&		&		\\
\hline
Cd	&3/4	&$-416$  	&$\uparrow\,\,0$, $\downarrow\,\,1.66$	&$4.39$	&$3.98$	&$235$ \\
	&1/4	&$578$	&$0.90$						&$-6.57$	&		&		\\
\hline
Hg	&3/4	&$-400$	&$\uparrow\,\,0$, $\downarrow\,\,0.97$	&$4.18$	&$3.99$	&$220$ \\
	&1/4	&$592$	&$0.87$						&$-7.21$	&		&		\\
\end{tabular}
\end{ruledtabular}
\end{table*}

\clearpage

\clearpage

\begin{figure}
\includegraphics[width=\textwidth]{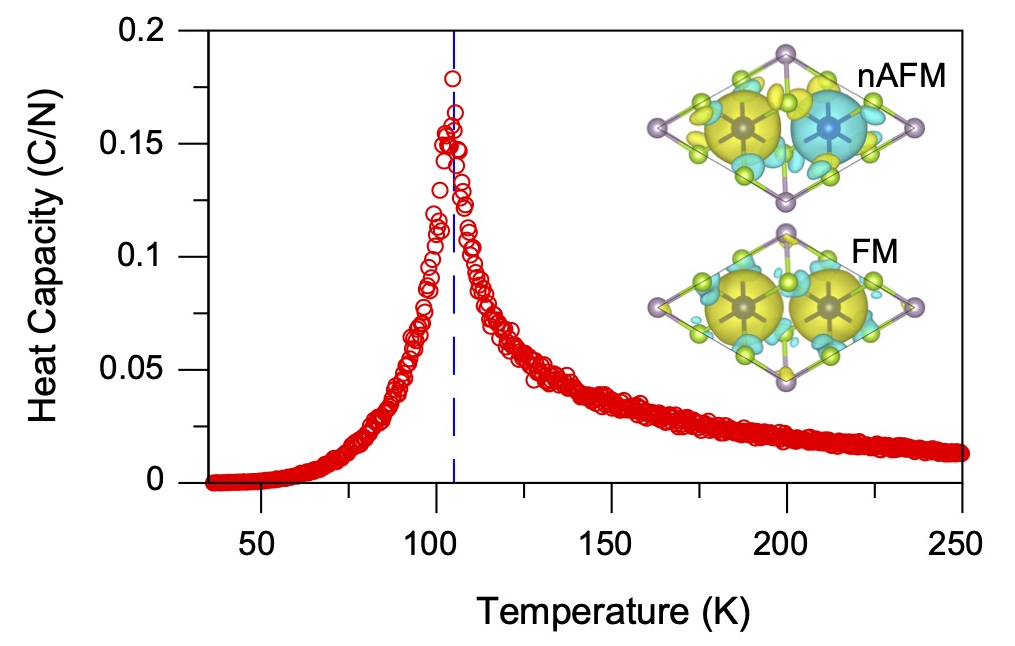}
\noindent Fig.\,S2. The Monte-Carlo simulated specific heat capacity as a function of temperature for 2D CrPSe$_3$.
\end{figure}

\clearpage

\begin{figure}
\includegraphics[width=\textwidth]{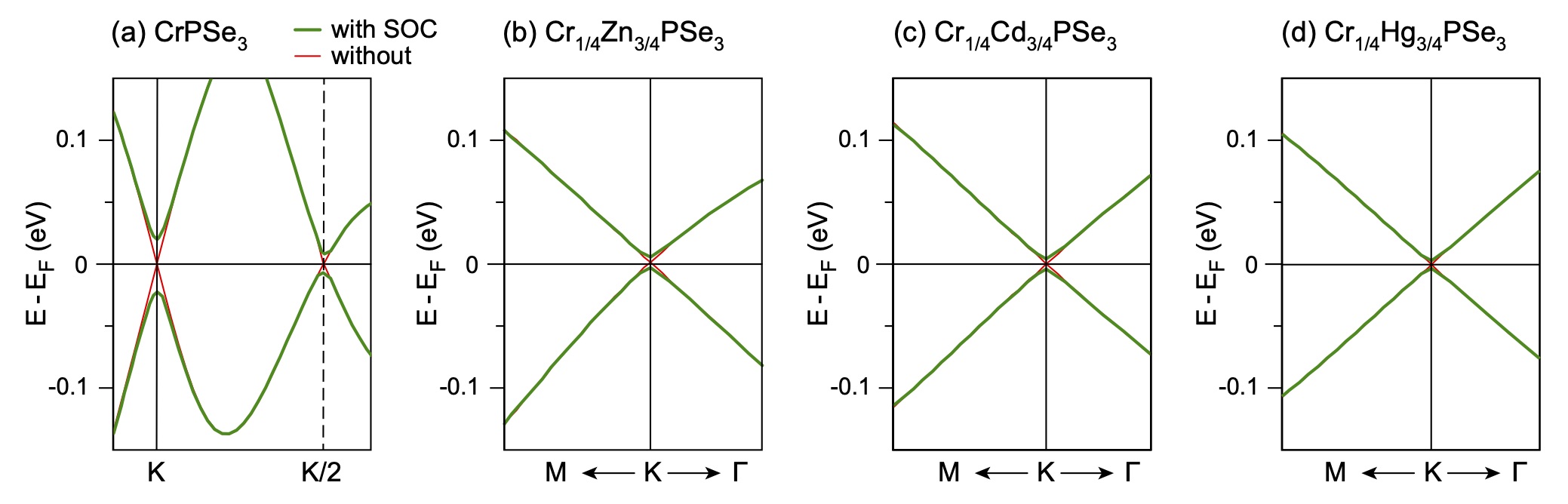}
\noindent Fig.\,S3. Comparison of band structures of FM-CrPSe$_3$ as well as mixed Cr$_{1/4}$M$_{3/4}$PSe$_3$ monolayers in the vicinity of $E_F$ obtained with DFT$+U$ and with DFT$+U$+SOC. 
\end{figure}

\clearpage

\begin{figure}
\includegraphics[width=\textwidth]{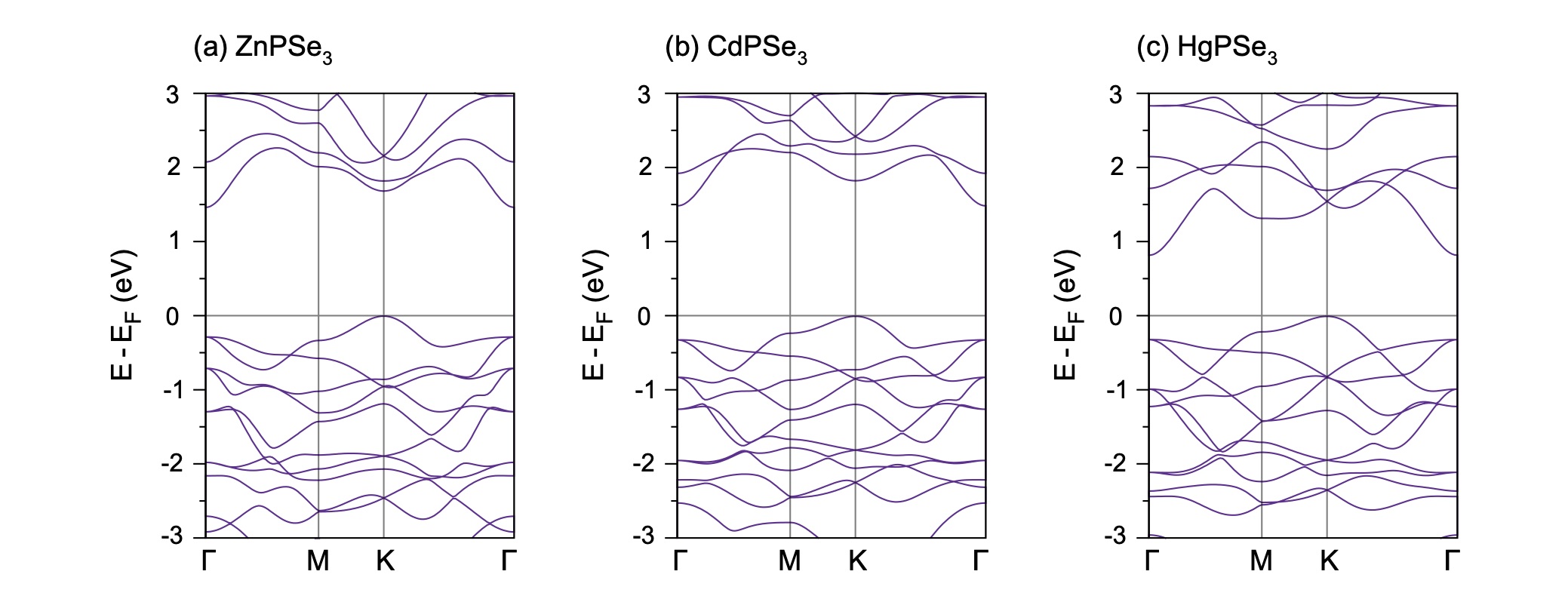}
\noindent Fig.\,S4. Band structures of 2D MPSe$_3$ (M = Zn, Cd, Hg).
\end{figure}

\clearpage

\begin{figure}
\includegraphics[width=\textwidth]{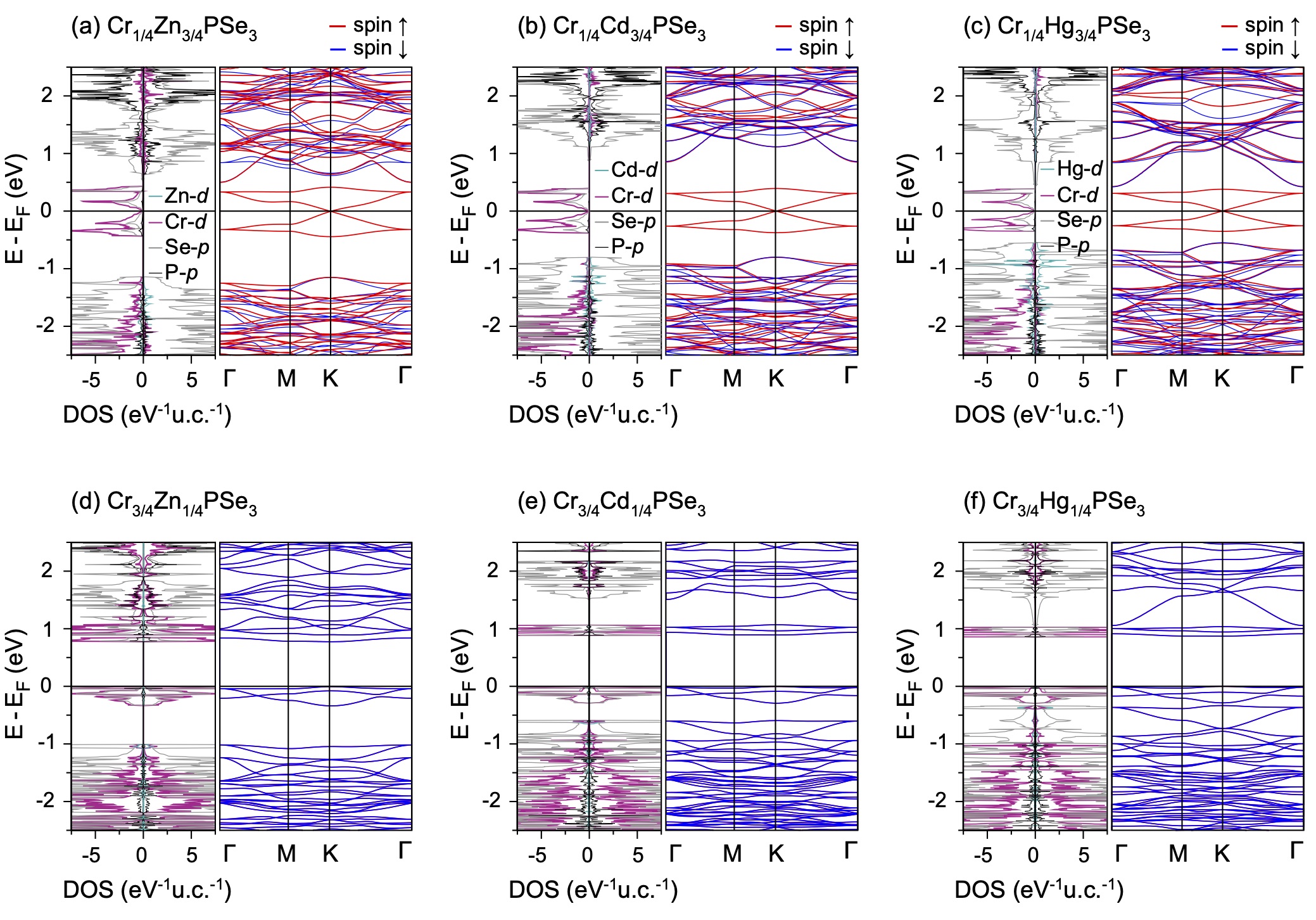}
\noindent Fig.\,S5. Band structures and orbital-projected density of states calculated for 2D Cr$_{1-x}$M$_{x}$PSe$_3$ (M = Zn, Cd, Hg; $x = 1/4, 3/4$) monolayers in their ground-state configurations.
\end{figure}

\clearpage

\begin{figure}
\includegraphics[width=\textwidth]{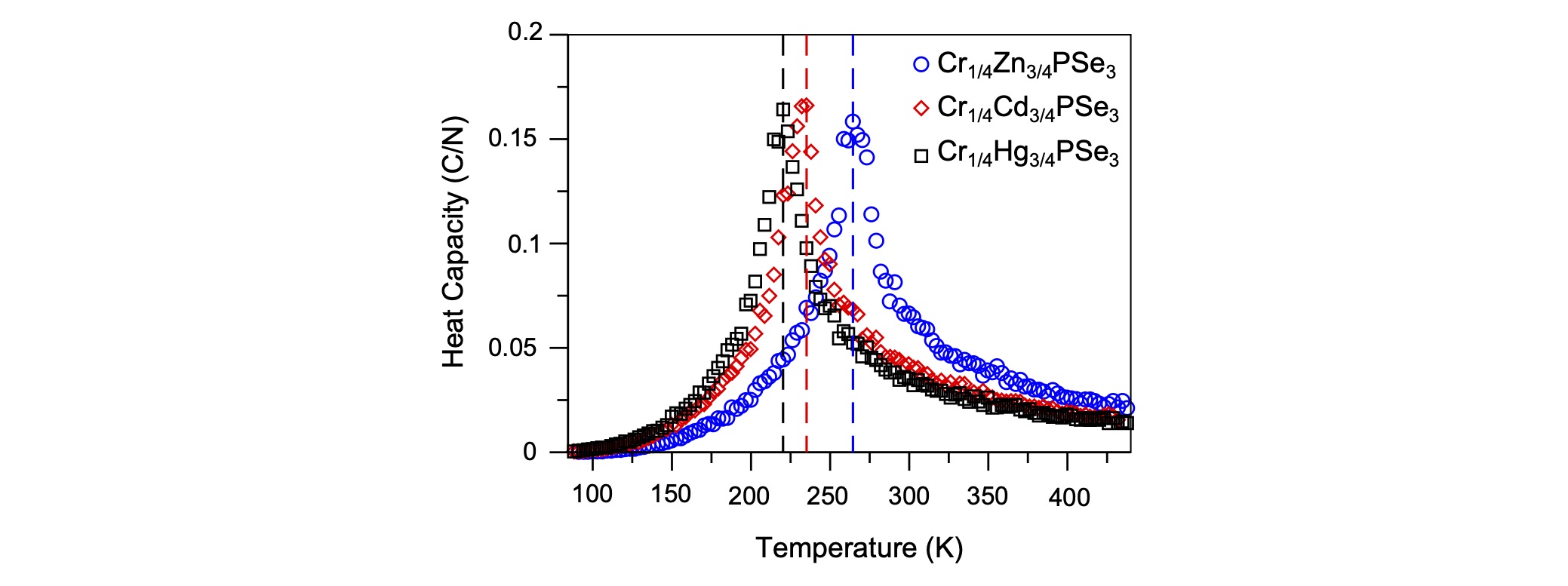}
\noindent Fig.\,S6. The Monte-Carlo simulated specific heat capacity as a function of temperature for 2D Cr$_{1/4}$M$_{3/4}$PSe$_3$ (M = Zn, Cd, Hg). 
\end{figure}

\end{document}